\documentstyle[prd,aps,floats,graphicx]{revtex}

\newcommand\gev{\mathrm{GeV}}
\newcommand\mev{\mathrm{MeV}}
\newcommand\mpl{M_{\mathrm{Pl}}}
\newcommand\req[1]{Eq.~(\ref{#1})}

\begin{document}
\draft

\twocolumn[\hsize\textwidth\columnwidth\hsize\csname
@twocolumnfalse\endcsname

\title{Quintessence arising from exponential potentials}
\author{T. Barreiro, E. J. Copeland and N. J. Nunes}
\address{Centre for Theoretical Physics,
University of Sussex, Falmer, Brighton BN1 9QJ,~~~U.~K. }

\date{\today}
\maketitle
\begin{abstract}
We demonstrate how the properties of the attractor solutions of
exponential potentials
can lead to models of quintessence with the currently 
observationally favored equation of state.
Moreover, we 
show that these properties hold for a wide range of initial 
conditions and for natural values of model parameters. 
\end{abstract}
\pacs{PACS numbers: 98.80.Cq \hspace*{1.5cm} SUSX-TH-99-016
\hspace*{1.5cm} astro-ph/9910214 }
\vskip2pc]


\section{Introduction}
Measurements of the redshift-luminosity distance relation
using high redshift type Ia supernovae combined with 
cosmic microwave background (CMB)
and galaxy clusters data appear to
suggest that the present Universe is flat and undergoing a
period of $\Lambda$ driven inflation,
with the energy density split into two main contributions,
$\Omega_{\rm matter} \approx 1/3$ and
$\Omega_{\Lambda}
\approx 2/3$ \cite{bahcall,riess,scp}.
Such a startling finding has naturally led theorists to
propose explanations for such a phenomenon. One such
possibility that has attracted a great deal of attention is the
suggestion that a minimally coupled homogeneous
scalar field $Q$ (the ``quintessence'' field),
slowly rolling down its potential,
could provide the dominant contribution to the energy density
today thanks to the special form of the
potential \cite{{caldwell},{zlatev}}.
Non-minimally coupled models have also been investigated
\cite{{uzan},{chiba},{amendola},{perrota},{holden},{bartolo}}.
The advantage of considering a more general component
that evolves in time so as to dominate the energy density
today, as opposed to simply inserting the
familiar cosmological constant is that the latter would
require a term $\rho_{\Lambda} \approx 10^{-47}
~\gev^4$ to be present at all epochs, a rather small value
when compared to typical particle physics scales. On the
other hand, quintessence models possess
attractor solutions which allow for a wide range of initial
conditions, all of which can correspond to the same energy
density today simply by tuning one overall multiplicative
parameter in the potential.

There is a long history to the study of scalar field
cosmology especially related to time varying cosmological
constants. Some of the most influential early
work is to be found in Refs.
\cite{{ratra},{peebles},{wetterich}}.
One particular case which at first sight appears promising
is the one involving exponential potentials of the form
$V \propto \exp(\lambda \kappa Q)$,
where $\kappa^2 = 8 \pi G$
\cite{{ratra},{peebles},{wetterich},{spokoiny},{wands},{ferreira},{copeland},{liddle}}.
These have two possible late-time attractors in
the presence
of a barotropic fluid: a scaling regime where the
scalar field mimics the dynamics of the background
fluid present, with a constant ratio between both energy
densities, or a solution dominated by the scalar field. The former
regime cannot explain the observed values for the
cosmological parameters discussed above; basically
it does not allow for an accelerating expansion
in the presence of a matter background fluid.
However, the latter regime does not
provide a feasible scenario either, as there is a tight
constraint on the allowed magnitude of  $\Omega_{Q}$ at
nucleosynthesis\cite{ferreira,copeland}. It turns
out that it must satisfy $\Omega_{Q}(1 \mev) < 0.13$. On the
other hand, we must allow time for formation of structure
before the Universe starts accelerating.
For this scenario to be possible we would have to fine tune
the initial value of $\rho_Q$, but this is precisely the kind
of thing we want to avoid.

A number of authors have proposed potentials which will lead
to $\Lambda$ dominance today.
The initial suggestion was an inverse power law potential
(``tracker type'') $V \propto Q^{- \alpha}$
\cite{{zlatev},{ratra},{liddle}}, which can be
found in models of supersymmetric
QCD \cite{{binetruy},{masiero}}. Here the ratio of energy
densities is no longer a constant but $\rho_{Q}$ scales
slower than $\rho_{B}$ (the background energy density) and
will eventually dominate. This epoch can be set conveniently
to be today by tuning the value of only one parameter in the
potential. However, although appealing,
these models suffer in that their predicted
equation of state $w_{Q}=p_{Q}/\rho_{Q}$ is marginally
compatible with the favored values emerging from
observations using SNIa and CMB measurements, considering a
flat universe \cite{{perlmutter},{wang},{efstathiou}}. For
example, at the 2$\sigma$ confidence level in the
$\Omega_M-w_Q$ plane, the data prefer $w_Q <
-0.6$ with a favored cosmological constant $w_Q = -1$
(see e.g. \cite{efstathiou}), whereas the values permitted by
these tracker potentials (without fine-tuning) have $w_Q > -0.7$
\cite{steinhardt}. For an interpretation of the data which allows
for $w_Q < -1$ see Ref.\cite{caldwell1}.

Since this initial proposal, a number of authors have made
suggestions as
to the form the quintessence potential could take
\cite{{vilenkin},{brax},{bento},{albrecht},{matos},{macorra},{sahni}}.
In particular, Brax and Martin \cite{brax} constructed a simple
positive scalar potential motivated from supergravity models, $V
\propto \exp(Q^2)/Q^{\alpha}$, and showed that even with the
condition $\alpha \geq 11$,
the equation of state could be pushed 
to $w_Q \approx -0.82$, for $\Omega_Q = 0.7$.
A different approach was followed by
the authors of \cite{{albrecht},{sahni}}. They investigated a
class of scalar field potentials  where the quintessence field
scales  through an exponential regime until it gets
trapped in a minimum with a non-zero vacuum energy, leading to a 
period of de Sitter inflation with $w_Q \to -1$.

In this Brief Report we investigate a simple class of potentials which
lead to striking results.
Despite previous claims, exponential potentials by themselves
are a promising 
fundamental tool to build quintessence potentials. 
In particular, we show that potentials
consisting of sums of exponential terms can easily deliver
acceptable models of quintessence in close agreement with
observations for natural values of parameters.

\section{Model}

We first recall some of the results presented in
\cite{{wetterich},{ferreira},{copeland}}.
Consider the dynamics of a scalar field $Q$, with an
exponential potential $V \propto \exp(\lambda \kappa Q)$.
 The field is evolving in a spatially flat Friedmann-Robertson-Walker
(FRW) universe with
a background fluid which has an equation
of state $p_B=w_B\rho_B$.  There exists just two possible
 late time attractor solutions with quite different
properties, depending on the values of $\lambda$ and
$w_B$:
 
(1) $\lambda^2 > 3(w_B + 1)$. The late time attractor
 is one where
the scalar field mimics the evolution of the barotropic fluid
 with $w_{Q} = w_B$, and the relation
$\Omega_{Q} = 3(w_B + 1)/\lambda^2$ holds. 

(2) $\lambda^2 < 3(w_B + 1)$. The late time attractor is the
scalar field dominated
solution ($\Omega_Q =1$) with $w_{Q} = -1 + \lambda^2/3$.

Given that single exponential terms can lead to one of the
above scaling solutions, then it
should follow that a combination of the above regimes should
allow for a scenario where the
universe can evolve through a radiation-matter regime
(attractor 1) and at some recent epoch
evolve into the scalar field dominated regime (attractor 2).
We will show that this does in fact occur
for a wide range of initial conditions. To provide a concrete
 example consider the following potential for a scalar field
$Q$:
\begin{equation}
\label{pot}
V(Q) = M^4( e^{\alpha \kappa Q} + e^{\beta \kappa Q} ),
\end{equation}
where we assume $\alpha$ to be positive
(the case $\alpha < 0$ can always be obtained
taking $Q \to - Q$).
We also require $\alpha >  5.5$, a constraint
 coming from the nucleosynthesis bounds on
$\Omega_Q$ mentioned earlier \cite{{ferreira},{copeland}}.

\begin{figure}[ht]
\includegraphics[height=6cm,width=8cm]{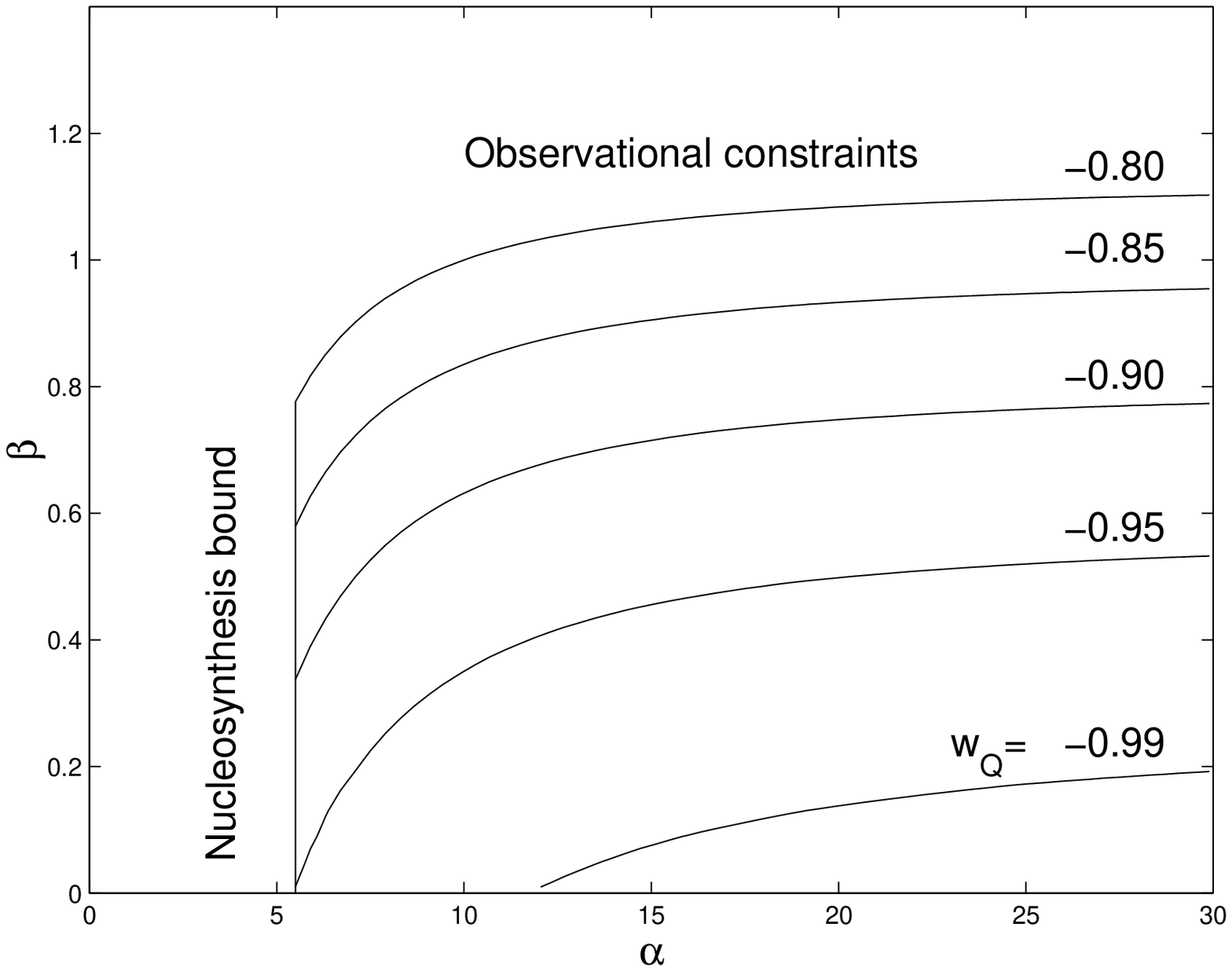}
\caption{}\label{param}
Contour plot of $w_Q({\mathrm{today}})$ as a function
of $(\alpha, \beta)$, with the constraint
$\Omega_Q(\mathrm{today}) \approx 0.7$.
The region $\alpha < 5.5$ is excluded because of the
nucleosynthesis bound, $\Omega_Q(1 \mev) < 0.13$, and the
upper region due to $1 \sigma$ observational constraints.
\end{figure}

First, we assume that  $\beta$ is also positive. In order to
have an idea of what the value of $\beta$ should be,
note that if today we  were in the regime dominated by the scalar field
(i.e.  attractor 2), then in order to satisfy observational
constraints for the quintessence equation of
state (i.e. $w_Q < -0.8$), we must have $\beta < 0.8$.
We are not obviously in the dominant regime today but
in the transition between the two regimes so this is just a
central value to
be considered. In Fig.~\ref{param} we show that acceptable
solutions to Einstein's equations
in the presence of radiation, matter and the quintessence
field can be accommodated
for a large range of parameters ($\alpha,~\beta$).

The value of $M$ in Eq.~(\ref{pot}) is chosen so that today
$\rho_Q \approx \rho_c \approx 10^{-47} \gev^4$. This then implies 
$M \approx 10^{-31} \mpl \approx 10^{-3} \mathrm{eV}$.
However, note that if we generalize the potential in \req{pot} to
\begin{equation}
\label{finepot}
V(Q) = \mpl^4( e^{\alpha \kappa (Q-A)} + 
                      e^{\beta \kappa (Q-B)} )\;,
\end{equation}
then all the parameters become of the order of the
Planck scale. Since the scaling regime of exponential
potentials does 
not depend upon its mass scale [i.e. $M$ in \req{pot}], $A$ is actually
a free parameter
that can, for simplicity, be set to $\mpl$ or even to zero.
On the other hand, just like before for $M$, $B$
needs to be such that today we obtain the right value 
of $\rho_Q $. 
In other words, we require
$M^4 \sim \mpl^4 e^{-\beta B} \sim \rho_Q $.
This turns out to be $B = {\cal O}(100) \mpl$, 
depending on the precise values of $\alpha$, $\beta$ and $A$.

\begin{figure}[ht]
\includegraphics[height=6cm,width=8cm]{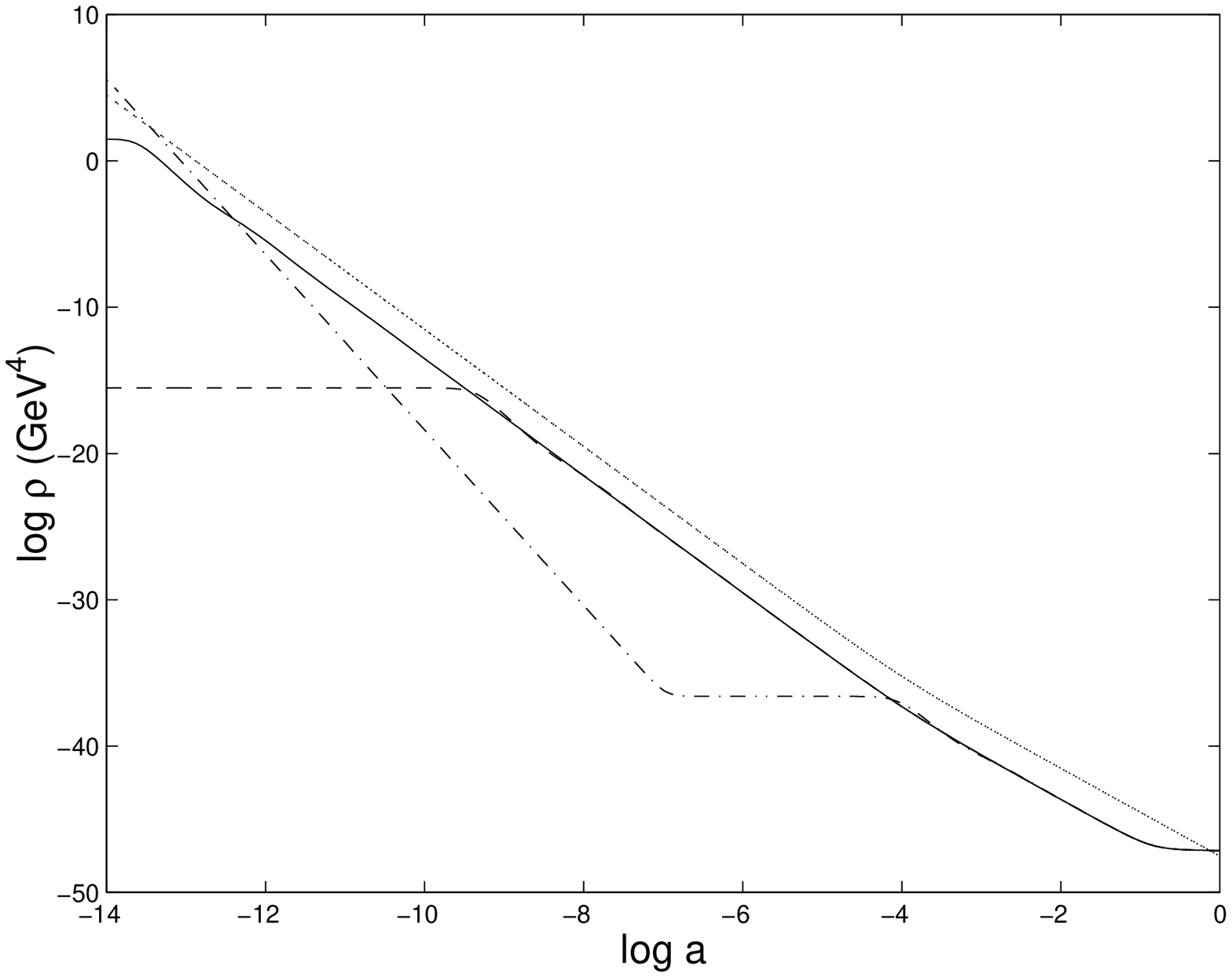}
\caption{}\label{endens} Plot of the energy density, $\rho_Q$,
 for $\alpha = 20$, $\beta = 0.5$ and several initial
conditions admitting an $\Omega_Q = 0.7$
flat universe today. The solid line represents the evolution
which emerges from equipartition at the end of inflation
and the dotted line represents $\rho_{ \rm matter}+ \rho_{\rm radiation}$.
\end{figure}

There is another important advantage to the potentials of
the form in Eq.~(\ref{pot}) or Eq.~(\ref{finepot}); 
namely, we obtain
acceptable solutions for a wider
range of initial energy densities of the quintessence field 
than we would with say the inverse power law potentials. 
For example, 
in Fig.~\ref{endens} we show that it is perfectly
acceptable to start with the energy density of the 
quintessence field above that of radiation, and still enter 
into a subdominant scaling regime at later times; however, 
this is an impossible feature in the context of inverse 
power law type potentials \cite{steinhardt}.

Another manifestation of this wider class of solutions can be 
seen by considering the case where the field evolution began 
at the end of an initial period of inflation. In that case,
as discussed in Ref.\cite{steinhardt}, we
could expect that the energy density of the system is equally 
divided among all the
thousands of degrees of freedom in the cosmological fluid.
This equipartition of energy would imply that just after
inflation $\Omega_i \approx 10^{-3}$.
If this were the case, for inverse
power law potentials, the power could not be smaller
than $5$ if the field was to reach the attractor by 
matter domination.
 Otherwise, $Q$ would freeze at some value and simply act as
a cosmological constant until the present (a perfectly
acceptable scenario of course,
but not as interesting). Such a bound on the power implies
 $w_Q > -0.44$ for $\Omega_Q = 0.7$.
With an exponential term,
this constraint is considerably weakened.
Using the fact that the field is frozen at a value
$Q_f \approx Q_i - \sqrt{6\, \Omega_i}/\kappa$, where
$Q_i$ is the initial value of the field \cite{steinhardt}, we can
see that the equivalent problem only arises when
\begin{equation}
\alpha \sqrt{6 \Omega_i} - 2 \ln \alpha \gtrsim
\ln \left ( \frac{\rho_{Q_i}}{2 \rho_{eq}} \right ),
\end{equation}
where $\rho_{Q_i}$ is the initial energy density of the
scalar field
and $\rho_{eq}$ is the background
energy density at radiation-matter equality.
For instance, for our plots
with $a_i = 10^{-14}$, $a_{eq} = 10^{-4}$,
this results in a bound $\alpha \lesssim   10^3$.

A new feature arises when we consider potentials
of the form given in Eq.~(\ref{pot}) with the 
 nucleosynthesis bound $\alpha > 5.5$ 
but taking this time $\beta < 0$.
In this case the potential has a minimum at
$\kappa \, Q_{\rm min} =  \ln ( - \beta / \alpha)/(\alpha - \beta)$
with a corresponding
value $V_{\rm min} = M^4 \frac{\beta - \alpha}{\beta}
(-\frac{\beta}{\alpha})^{\alpha/(\alpha - \beta)}$.

\begin{figure}[ht]
\includegraphics[height=6cm,width=8cm]{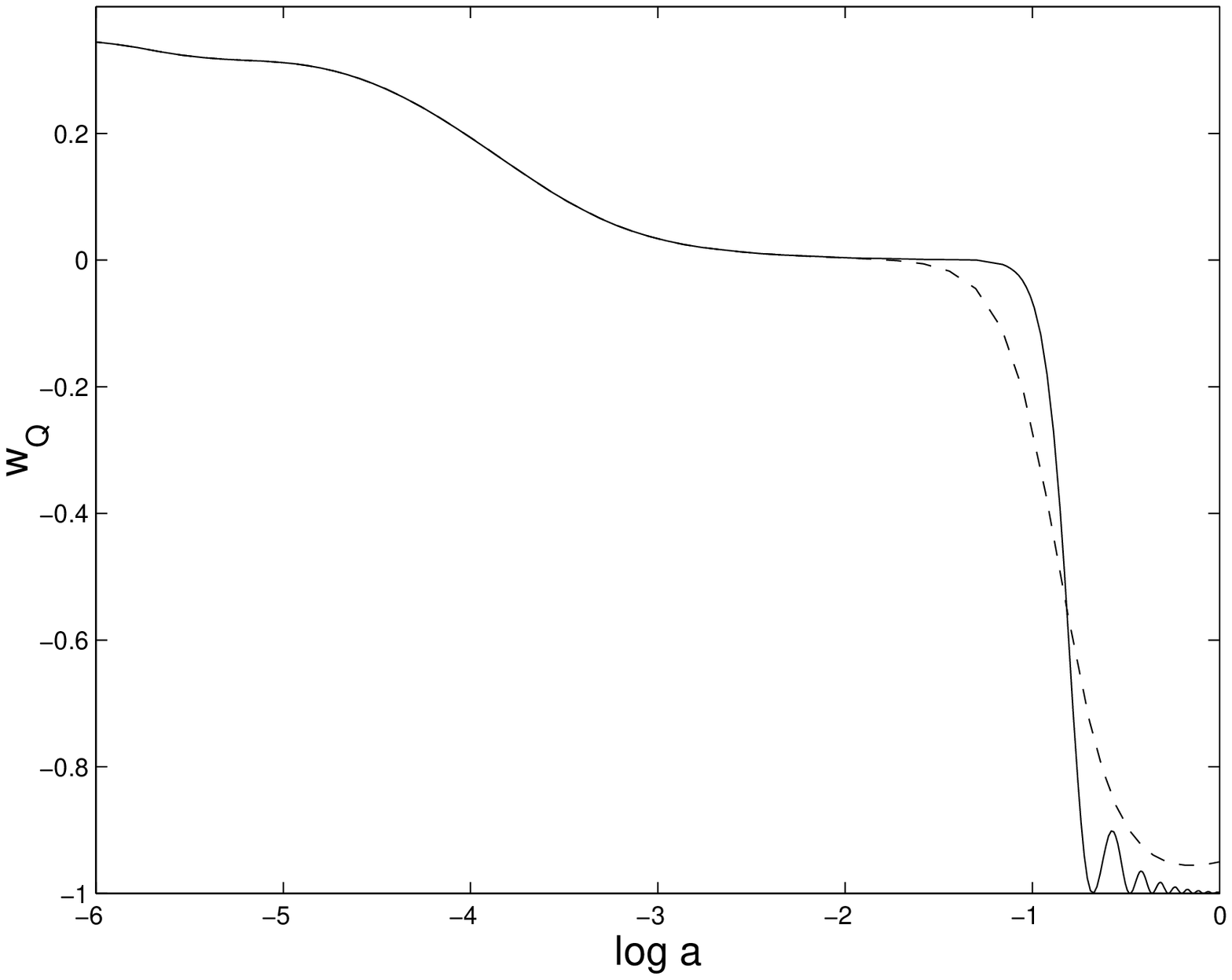}
\caption{}\label{eqstate} The late time evolution of the
equation of state for
parameters $(\alpha,\beta):$ dashed line (20,0.5); solid
line (20, $-$20)
and $\Omega_Q \approx 0.7$. ($a_0 = 1$ today).
\end{figure}

Far from the minimum, the scalar field scales as described above
(attractor 1). However, when the field reaches the minimum, the
effective cosmological constant $V_{\rm min}$ will quickly take
over the evolution as the oscillations are damped, driving the
equation of state towards $w_Q = -1$. This scenario is illustrated
in Fig.~\ref{eqstate}, where the evolution of the equation of
state is shown and compared to the previous case with $\beta >
0$. In many ways this is the key result of the paper, as in this
figure it is clearly seen that the field scales the radiation ($w
= 1/3$) and matter ($w = 0$) evolutions before settling in an
accelerating ($w < 0$) expansion. Once again, as a result of the scaling
behavior of attractor 1, it is clear that there exists
a wide range of initial conditions that provide realistic results.
The feature resembles the recent suggestions of Albrecht and
Skordis \cite{albrecht}. The same
mechanism can be used to stabilize the dilaton in string theories
where the minimum of the potential is fine-tuned to be zero rather
than the non-zero value it has in these models \cite{{barreiro}}.

In \cite{steinhardt}, a quantity $\Gamma \equiv V'' V/(V')^2$ is
proposed as an indicator of how well a given model converges to a
tracker solution. If it remains nearly constant, then the solutions
can converge to a tracker solution. It is easy to see from
Eq.~(\ref{pot}) that apart from the transient regime where the
solution evolves from attractor 1 to attractor 2, $\Gamma =1$
to a high degree of accuracy.

It is important to note that for this mechanism
to work, we are not limited to potentials containing
only two exponential terms and one field.
Indeed, all we require of the dynamics is to enter
one period like regime 1,
which can either be followed by one regime like
2, or by the field settling in a minimum with a non-zero vacuum energy.
We can consider as an example the case
of a potential depending on two fields of the form
\begin{equation}
\label{pot2fields}
V(Q_1,Q_2) = M^4( e^{\alpha_1 \kappa Q_1 + \alpha_2 \kappa Q_2} +
e^{\beta_1 \kappa Q_1 + \beta_2 \kappa Q_2} ),
\end{equation}
where all the coeficients are positive. 
This leads to similar results to Eq.~(\ref{pot}) for a single 
field $Q$, with effective early and late slopes  given by  
$\alpha_{\rm eff}^2 = \alpha_1^2 +\alpha_2^2$ and   
$\beta_{\rm eff}^2 = \beta_1^2 + \beta_2^2$, respectively.
Such a result is not surprising and is caused by 
the assisted behavior that can    
occur for multiple fields \cite{assisted}. 
Note that for this type of multiple   
field examples  the effective slopes in the resulting 
effective potential are larger than the individual slopes, 
a useful feature since we require $\alpha_{\rm eff}$ to be large.   

\section{Discussion}
 So far, we have presented a series of potentials
that can lead to the type of quintessence behavior capable of
explaining the current data arising from high redshift type Ia
supernovas, CMB and cluster measurements. The beautiful
property of exponential potentials is that they lead to scaling
solutions which can either mimic the background fluid or dominate
the background dynamics depending on the slope of the potential.
We have used this to develop a picture where by simply
combining potentials of different slopes, it is easy to obtain
solutions which first enter a period of scaling through the 
radiation and matter
domination eras and then smoothly evolve to dominate the energy
density today. 
We have been able to demonstrate that the
quintessence behavior occurred for a wide range of initial
conditions of the field, whether $\rho_{Q}$ be initially 
higher or lower than $\rho_{\rm matter} + \rho_{\rm radiation}$.
We have also shown that the favored observational values for 
the equation of state $w_Q({\mathrm{today}}) < -0.8$
can be  easily reached for 
natural values of the parameters in the potential. 
This is a big improvement in respect to most quintessence 
models as they usually give either $w_Q \gtrsim -0.8$ or $w_Q = -1$.

We have to ask, how sensible are such
potentials? Can they be found in nature and, if so, can we make use
of them here? The answer to the first question seems to be, yes
they do arise in realistic particle physics models
\cite{stelle,reall,lu,lavrinenko,reall99,green}, but the current models do
not have the correct slopes. Unfortunately, the tight constraint
emerging from nucleosynthesis, namely $\alpha > 5.5$, is difficult
to satisfy in the models considered to date which generally have
$\alpha \leq 1$. It remains a challenge to see if such potentials
with the required slopes can arise out of particle physics.
One possibility is that the desirable slopes will be obtained from
the assisted behavior when several fields are present as
mentioned above.

It is encouraging that the quintessence behavior
required to match current observations occurs for such simple
potentials.

\acknowledgements
We would like to thank Orfeu Bertolami, Robert Caldwell,
Thomas Dent, Jackie Grant, Andrew Liddle, Jim
Lidsey and David Wands for useful
discussions. E.J.C. and T.B. are supported by PPARC. N.J.N. is
supported by
FCT (Portugal) under contract PRAXIS XXI BD/15736/98. E.J.C is
grateful to the staff of the
Isaac Newton Institute for their kind hospitality during the
period when part of this work was being completed.

\end{document}